\documentstyle[amssymb,aps,preprint]{revtex}

\begin{document}
\title{Finite-wavevector Jahn-Teller-pairing and superconductivity in the cuprates. }
\author{D.Mihailovic and V.V.Kabanov}
\address{Jozef Stefan Institute, Jamova 39, 1001 Ljubljana, Slovenia}
\maketitle

\begin{abstract}
A model interaction is proposed in which pairing is caused by a{\em \
non-local }Jahn-Teller (JT) -like instability due to the coupling between
planar O states and $k\neq 0$ phonons. Apart from pairing, the interaction
is found to naturally allow metallic stripe formation. The consequences of
the model for superconductivity in the cuprates are discussed. The model is
shown to be consistent with numerous sets of experimental data in quite some
detail.
\end{abstract}


\newpage

\section{Introduction}

A Jahn-Teller (JT) polaron pairing effect was originally proposed as a
possible explanation for the superconductivity in La$_{2-x}$Ba$_{x}$CuO$_{4}$
by Bednorz and M\"{u}ller\cite{BednorzMuller}. Since then the JT effect has
been discussed by a number of authors in different contexts\cite
{Gorkov,Weger,Markiewicz,Kresin,EgamiJT} and although many features have
been observed experimentally supporting the general concept of JT\ polarons 
\cite{Muller,GuoMeng}, so far no generally applicable model has been shown
to be compatible with the overall phenomenology observed in the cuprates.
One of the major problems is that the single-ion JT energy splitting between
Cu $d_{x^{2}-y^{2}}$ states and d$_{3r^{2}-z^{2}}$ states is thought to be
of the order of 1 eV or more, too large to play a role in the pseudogap
physics, which is believed to be the energy scale of the pairing
interaction, which is of the order of 0.1 eV. Nevertheless, the observation
of a large isotope effect on both $T_{c}$\cite{isotopeTc}, $T^{\ast }$\cite
{GuoMeng} and penetration depth\cite{pendep} firmly establishes a role for
lattice polarons in the pairing mechanism, while the fact that a depression
in the spin susceptibility usually appears at a lower temperature than the
''pseudogap'' observed by charge excitation spectroscopies\cite{Mook,NMR}
suggests that a lattice pairing mechanism is primary and the spin ordering
follows.

In this paper we outline a new type of microscopic pairing scenario in La$%
_{2-x}$Sr$_{x}$CuO$_{4}$ driven by a finite-wavevector JT instability. We
find that the proposed model can explain many of the general features both
in the underdoped and overdoped regions of the phase diagram and is
fundamentally compatible with the overall phenomenology of the cuprates.

The experimental observations on which the present scenario is based are
mainly those showing evidence for the existence of dynamic incommensurate
lattice distortions associated with doped holes. Inelastic neutron
scattering \cite{Mook,Egami}, neutron PDF\cite{Sendyka}, EXAFS\cite
{Bianconi,Bozin} and ESR\cite{Kochelaev} experiments all show the existence
of dynamic lattice distortions on timescales relevant for pairing of 10$%
^{-13}-$10$^{-15}$s. The inelastic neutron scattering data\cite{Egami,Mook}
can be singled out for {\em directly} giving not only the energy, but also
the wavevector associated with the lattice distortion and its range in $k$%
-space without any interpretation or modeling. The observed distorted
regions appear to be along the ($\zeta ,0,0$) (or ($0,\zeta ,0$))
directions, and have typical dimensions in real space of $2a\times 5a$ $%
(8\times 20$ \AA ), where $a$ is the lattice constant. A shematic diagram of
the distortion derived from an analysis of the data is shown in Figure 1.
The energy of the anomaly of $E_{a}=65\sim 85$ meV is of the order of the
''pseudogap'' energy, while its width of $\Delta E\simeq 5$ meV corresponds
closely to the linewidth expected from the measured pair recombination rate, 
$h/\pi c\tau \simeq 4$ meV\cite{Demsar}. At the doping level of $x=0.15$,
the 2D volume of the object in Figure 1 contains approximately 1.5 carriers.
Taken together, the implication is that the objects can be interpreted as $k$%
-space ''snapshots'' of individual pairs.

Additional experimental observations which we consider important in the
present context is evidence for the co-existence of two carrier types in a
large part of the phase diagram\cite{Mullersusc,Bled}, and - in addition to
the ''pseudogap'' - the appearance of a temperature-dependent
superconducting gap $\Delta _{c}(T)$ which closes at $T_{c}$\cite
{Norman,Demsar} which is particularly well observed at higher doping levels
and has a magnitude at $T=0$ of $\Delta _{c}(0)\lesssim \Delta _{p}$.

\section{JT\ pairs and stripes.}

Before proceeding with the analysis of the {\it e-p} coupling for the case
of general $k$, let us briefly discuss the $\Gamma -$ point coupling ($k=0$)
in the tetragonal group $D_{4h}^{17}$ (I4/mmm) applicable to La$_{2-x}$Sr$%
_{x}$CuO$_{4}.$ The symmetrised cross product of the representations at the $%
\Gamma $ point is 
\begin{equation}
\lbrack E_{u}\times E_{u}]=[E_{g}\times E_{g}]=A_{1g}+B_{1g}+B_{2g},
\end{equation}
which contains no degenerate representations. On the other hand, the lattice
vibrations at the $\Gamma -$point transform as: 
\[
\Gamma =2A_{1g}+4A_{2u}+B_{2u}+2E_{g}+5E_{u}. 
\]
Since there are no $B_{1g}$ and $B_{2g}$ representations at the $\Gamma $
point, electrons can couple only with $A_{1g}$ phonon modes. In $D_{4h}$
there are two such modes associated with apex oxygens or La ions. However,
the experiments show that the modes involved in the intraction are those of
in-plane O atoms, which do not couple at the $\Gamma $ point. This leads us
to the main conjecture of the proposed pairing model, namely the existence
of {\em intersite }pairs which form via a $k\neq 0$ interaction.

The existence of intersite pairs in cuprate superconductors is inferred from
their very short coherence length. Given that the pair dimensions $l_{p}$
cannot exceed the coherence length, i.e. $l_{p}\lesssim \xi $, we may infer
that any possible lattice distortions associated with pairing have a finite
range $\sim l_{p}.$ The effect of such lattice distortions should also be
evident in reciprocal space, with an anomaly centered around a wavevector $%
k\simeq 1/l_{p}$. Following the inelastic neutron scattering data \cite
{Egami} which shows an anomaly approximately at $k_{0}\simeq \lbrack \pm \pi
/2a,0,0]$ extending over almost half the Brillouin zone (BZ) $\Delta k\sim
1/2a,$ we can write the electron-phonon interaction for such an object in
the form: 
\begin{equation}
g(k_{0},k)=g_{0}/((k-k_{0})^{2}+\gamma ^{2})
\end{equation}
where $g_{0}$ is a constant describing the strength of coupling, $k_{0}\,$\
defines the wavevector associated with the interaction and its range in $k$%
-space, which - neglecting fluctuations - also defines its extent in real
space (inter-hole spacing) as $l_{p}\sim k_{0}^{-1}$. $\gamma =\Delta k$
defines its width in $k$-space and gives the width of the distribution of
inter-carrier distances within the interacting pair. This is related to the
average size of the deformation of each particle in real space $\gamma ^{-1}$%
.

We now proceed with an analysis of the $e-p$ coupling using group theory for 
$k\neq 0\,\ $intersite pairing and first discuss the relevant phonon modes.
The BZ corresponding to the tetragonal space group $D_{4h}^{17}$ applicable
for La$_{2-x}$Sr$_{x}$CuO$_{4}$ is shown in Figure 2. To consider local
pairs and/or stripes forming along the Cu-O bond direction or along 45$%
^{\circ }$ to it, we need to consider the general wavevector $\Sigma $ and
the $\Delta $ points, corresponding to the ($\zeta ,0,0)$ and ($\zeta ,\zeta
,0)$ directions respectively. (The special symmetry points ($\Gamma ,X$ and $%
M$ etc.) give rise to commensurate distortions which will be discussed
later.) The relevant lattice deformation associated with the neutron mode at
75 meV \cite{Egami} (Fig. 1) is of $\tau _{1}$ - symmetry, where $\tau _{1}$
is the irreducible representation of the little group corresponding to the $%
\Sigma $ direction in the BZ as shown in Figure 3. Since in principle all
modes of $\tau _{1}$ symmetry can couple to electrons, for completeness we
show all the possible modes with $\tau _{1}$ symmetry in Figure 3. However,
the most relevant mode - i.e. the one for which the anomaly is observed to
be most pronounced - involves in-plane O1 displacements along the Cu-O bonds
(see also Fig.1).

\subsection{$k\neq 0$ phonon coupling to non-degenerate electronic states$.$}

Since the $\Sigma $ point has a four pronged star in $D_{4h}$, the coupling
of electrons in single non-degenerate electronic states to $k\neq 0$ phonons
can be written as: 
\begin{equation}
H_{int}=\sum_{{\bf l},s}n_{{\bf l},s}\sum_{k_{0}=1}^{4}\sum_{{\bf k}}g(k_{0},%
{\bf k})\exp {(i{\bf kl})}(b_{-{\bf k}}^{\dagger }+b_{{\bf k}})
\end{equation}
where ${\bf l}$ is the site label, and 
\begin{equation}
g(k_{0},{\bf k})=g(\pi \gamma ^{2})^{1/2}/((k-k_{0})^{2}+\gamma ^{2})
\end{equation}
where $k_{0}$ are the 4 wavevectors corresponding to the prongs of the star
associated with the interaction. The nondegenerate electronic states in this
interaction allowed by symmetry are associated with $p_{z}$-orbitals of
planar oxygens, and transform as $A_{2u}$ or $B_{2u}$ representations of the 
$D_{4h}$ symmetry group. However, the Hamiltonian (3) above on its own does
not lead to symmetry breaking, and thus is not of direct relevance for pair
or stripe formation.

\subsection{$k\neq 0$ phonon coupling to degenerate electronic states
(Jahn-Teller-like pairing)$.$}

A more interesting case arises when two-fold degenerate levels (for example
the two $E_{u}$ states corresponding to the planar O $p_{x}$ and $p_{y}$
orbitals or the $E_{u}$ and $E_{g}$ states of the apical O) interact with $k%
\not=0$ phonons. We are particularly interested in the phonons which lead to
symmetry breaking and allow the formation of intersite pairs or stripes. In
Eq. (5) we give the invariant Hamiltonian which couples degenerate
electronic states to phonons transforming as the $\tau _{1}$ representations
of the group of wave-vector $G_{k}$. Taking into account that $E_{g}$ and $%
E_{u}$ representations are real and Pauli matrices $\sigma _{i}$
corresponding to the doublet of $E_{g}$ or $E_{u}$ transform as $A_{1g}$ ($%
k_{x}^{2}+k_{y}^{2}$) for $\sigma _{0}=\left( 
\begin{array}{cc}
1 & 0 \\ 
0 & 1
\end{array}
\right) $, $B_{1g}$ ($k_{x}^{2}-k_{y}^{2}$) for $\sigma _{3}=\left( 
\begin{array}{cc}
1 & 0 \\ 
0 & -1
\end{array}
\right) $, $B_{2g}$ ($k_{x}k_{y}$) for $\sigma _{1}=\left( 
\begin{array}{cc}
0 & 1 \\ 
1 & 0
\end{array}
\right) $, and $A_{2g}$ ($s_{z}$) \ for $\sigma _{2}=\left( 
\begin{array}{cc}
0 & -i \\ 
i & 0
\end{array}
\right) $representations respectively, an invariant Hamiltonian is given by: 
\begin{eqnarray}
H_{int} &=&\sum_{{\bf l},s}\sigma _{0,{\bf l}}\sum_{k_{0}=1}^{4}\sum_{{\bf k}%
}g_{0}(k_{0},{\bf k})\exp {(i{\bf kl})}(b_{-{\bf k}}^{\dagger }+b_{{\bf k}})+
\nonumber \\
&&\sum_{{\bf l},s}\sigma _{3,{\bf l}}\sum_{k_{0}=1}^{4}\sum_{{\bf k}%
}g_{1}(k_{0},{\bf k})(k_{x}^{2}-k_{y}^{2})\exp {(i{\bf kl})}(b_{-{\bf k}%
}^{\dagger }+b_{{\bf k}})+ \\
&&\sum_{{\bf l},s}\sigma _{1,{\bf l}}\sum_{k_{0}=1}^{4}\sum_{{\bf k}%
}g_{2}(k_{0},{\bf k})k_{x}k_{y}\exp {(i{\bf kl})}(b_{-{\bf k}}^{\dagger }+b_{%
{\bf k}})+  \nonumber \\
&&\sum_{{\bf l},s}\sigma _{2,{\bf l}}S_{z,{\bf l}}\sum_{k_{0}=1}^{4}\sum_{%
{\bf k}}g_{3}(k_{0},{\bf k})k_{0}^{2}\exp {(i{\bf kl})}(b_{-{\bf k}%
}^{\dagger }+b_{{\bf k}})  \nonumber
\end{eqnarray}
where 
\begin{equation}
g_{i}(k_{0},{\bf k})=g_{i}(\pi \gamma ^{2})^{1/2}/((k-k_{0})^{2}+\gamma ^{2})
\end{equation}

The first term in (5) describes symmetric coupling and is identical to the
non-degenerate case (Eq. (3)). The second and third terms describe the e-p
interaction corresponding to the $\Sigma $ and $\Delta $ directions
respectively, while the last term describes the {\em coupling to spins}\cite
{Khomskii}.

The proposed interaction (5) on its own results in a splitting of the
degenerate states, breaking the tetragonal symmetry and resulting in a local
orthorhombic distortion at $k_{0}$ extending over $\gamma $ in $k$-space. It
can therefore lead to the formation of bound intersite pairs and/or stripes
with no further interactions. Of course the stability and size of such a
distortion will be determined by the balance of short-range attraction,
long-range Coulomb repulsion and kinetic energy \cite{Kuzmartsev}.

Now let us discuss the properties of the system governed by this Hamiltonian
(Eq. (5)). The importance of the different terms is of course to be
determined by experiments. For example, the large 20\% anomaly in inelastic
neutron scattering at the $\Sigma $ point clearly emphasizes the second ($d$%
-wave) term, while the absence of strong anomalies at $k=0$ de-emphasizes
the symmetric ($s$-wave) term and so the new ground state is expected to be
a pair which extends over a few unit cells along the Cu-O bond direction $%
(\pm \zeta ,0,0)$ or $(0,\pm \zeta ,0)$.\cite{direction} The internal
lattice structure within the pair is distorted, so pairing would be
associated with a reduced {\it local} symmetry within the pair. In other
words, the tetragonal or pseudo-tetragonal symmetry of the crystal is broken
locally by the formation of a non-local JT pair with a binding energy $%
E_{JT} $ given by the solution to the Hamiltonian (5). We thus associate the
pairing energy gap $E_{JT}$ with the experimental observation of a
''pseudogap'' at $kT^{\ast }\sim \Delta _{p}=E_{JT}$\cite{NMR,Demsar}.

To understand these finite-wavevector JT pairs in the context of the phase
diagram of the cuprates, we consider the effect of thermal fluctuations as
the temperature is reduced through $T^{\ast }$ in the underdoped phase (Fig.
4). For $T>T^{\ast }$ thermal energy prevents the carriers from forming
pairs at all levels of doping (shown schematically in the top row in Figure
5). Approaching $T^{\ast },$ JT\ pairs start to form and exist in
equilibrium with unbound carriers according to chemical balance at
thermodynamic equilibrium$\ n_{unbound}\sim \exp [-E_{JT}/k_{B}T],$ and
shown schematically in the lower panel of Fig. 5b). (The doping dependence
of $\Delta _{p}$ which is observed to approximately follow an inverse law $%
\Delta _{p}\sim 1/x$ \cite{Mullersusc,Demsar,Kabanov} is suggested to be a
result of screening as discussed by Alexandrov, Kabanov and Mott\cite{AKM},
and will not be discussed further here.)

For such non-local pairs to be stable, the energy gained by the JT pairing
must counteract the Coulomb repulsion between two charge carriers within the
pair $E_{JT}\gtrsim V_{i}$. The {\it upper limit }for the Coulumb repulsion
between two carriers approximately one coherence length apart is given by $%
V_{i}=e^{2}/4\pi \varepsilon r$ $\simeq 0.15$ eV (taking $r=1/k_{0}\simeq 2$
nm $\lesssim \xi _{s}$ and $\varepsilon =4$\cite{Timusk}). However, since $%
\varepsilon (\omega )\gg 4$ in the relevant frequency range for pairing (1 -
4 THz)\cite{Timusk}, the relevant value of $V_{i}$ can be significantly
smaller and can be easily overcome by $E_{JT}$.

Once pre-formed bosonic pairs exist, superconductivity can occur when phase
fluctuations between these pairs are sufficiently reduced so that phase
coherence can be established between them. This can occur by Bose
condensation \cite{Alexandrov,Uemura,BEC} or some form of the
Kosterlitz-Thouless transition \cite{Pokrovsky,EK}. In both cases the
critical transition temperature in the underdoped region of the phase
diagram is given by an expression relating $T_{c}$ to the pair density $%
n_{p} $ \ and effective mass $m^{\ast }$: 
\begin{equation}
T_{c}\simeq \hbar ^{2}n_{p}^{2/D}/(2m^{\ast })k_{B}
\end{equation}
where $D$\ is the dimensionality of the system\cite{interactions}. An
important issue related to whether Bose condensation occurs or another
mechanism is responsible for the formation of the condensate formation is
the number of pairs per coherence volume $V_{\xi }$. Using the experimental
upper limit of coherence length in La$_{1.85}$S$_{0.15}$CuO$_{4}$ at $T=0$
of $\xi \simeq 20$ \AA\ and assuming for the moment a uniform carrier
density in the Cu-O planes, then at a carrier concentration of $x=0.15$
there are approximately 1.5 {\em pairs per coherence volume}. Considering
the experimental error and uncertainties in geometrical factors involved in
determining $\xi $, it is clear that a crossover seems to occur near optimum
doping from $n_{p}<1$ per $V_{\xi }$ to $n_{p}>1$ per $V_{\xi },$ implying a
crossover from a Bose-condensation to overlapping-pair superconductivity
scenario\cite{Uemura}. Importantly, both are consistent with the
finite-wavevector non-local JT-pairing interaction described here. (The
detailed mechanism for the formation of a phase-coherent condensate is not
the subject of the present paper and will not be discussed further here.)

\subsection{Stripes}

So far, the discussion concerned a intersite pairing-JT effect with two
particles involved. If more than two{\it \ }particles are involved in the
interaction, the effect of (5) is similar and provided $k_{0}>\gamma $ (Eq.
1), a JT\ distortion can occur along a stripe, for example. The internal 
{\em lattice }structure of such stripes is defined by the JT\ lattice
distortion, just as for pairs. The shape of these objects is determined
primarily by minimisation of the Coulomb energy, and the formation of 1D
stripes is clearly more favourable than 2D clusters in this respect\cite
{Kuzmartsev}. The incommensurability of the dynamic JT distortion given by $%
k_{0}$ means that the number of sites in the stripe is larger than the
number of carriers, resulting in a partially filled ground state. The
electronic wavefunction inside such stripes is {\em extended}, that is, it
extends throughout the entire stripe, and the macroscopic transport
properties in the normal state are thus expected to be dominated by hopping
or tunneling of carries {\it between} the stripes, rather than within them.
The elementary excitations of such objects are expected to be Fermionic and
metallic in character, which makes their statistics different than for the
JT pairs, which are Bosons. The JT\ stripes are expected to {\em coexist }%
both with JT pairs {\em and} unbound particles, with their relative
populations determined by chemical balance and the pair binding energy $%
E_{JT}$ compared to the stripe formation energy. A schematic real-space
''snapshot'' picture of this phase is shown in Figure 5c). Note that because
we have a four-pronged star for the $\Sigma -$point distortions, four
different types of stripes can form, each corresponding to one of the four $%
k_{0}$. Since the little group at the\ $\Sigma -$point does not have
inversion symmetry, the stripes can have a local polarisation (i.e have a
ferroelectric phase). This may explain the presence of a spontaneous
polarisation in these materials and the appearance of a pyroelectric effect
in La$_{2-x}$Sr$_{x}$CuO$_{4}$\cite{Pyro1} and other cuprates\cite{Pyro2}.

A most simple and appealing possibility is that superconductivity in the
presence of stripes still occurs via the same pre-formed pair scenario as
discussed in the previous section. However, the stripes then appear to have
a detrimental effect on superconductivity, because they take up carriers and
thus reduce the number of pairs.

\section{Overdoped regime}

As the density of doped holes increases with increasing doping, the spacing
between them becomes comparable to the pair size and they start to overlap,
so interactions between the pairs and stripes become important, and some
kind of collective or cooperative effect which extends over both types of
objects needs to be considered.

The Hamiltonian in Eq. (5) introduces a number of length scales (see Fig.
5c)). The first is the mean distance between the charge carriers in the pair
(or within the stripe) $l_{p}\simeq 1/k_{0}$. The second is the length of
the stripes $l_{s}$ and finally, there is the length scale $l_{c}$
describing the characteristic distance {\em between} the pairs or stripes,
which is determined simply by the carrier density.

With increasing doping, the distance between the pairs and stripes $l_{c}$
decreases and increased screening reduces Coulomb repulsion, which in turn
leads to increased stripe length $l_{s}$. At some point $l_{c}$ becomes
comparable to the superconducting coherence length $\xi _{s}$, and the
superconducting pairs become {\em proximity coupled} to the metallic stripes
(Fig. 5c). In other words, superconductivity in the stripes will be induced
below $T_{c}$ by a JT pair-gap proximity effect. Above $T_{c}$, there is no
proximity coupling and so clearly the superconducting order parameter must
be zero in the stripes. Thus it is evident that the superconducting order
parameter has to be $T$-dependent within the stripes. Whence an explanation
for the experimentally observed {\em coexistence} of a $T$-independent
pairing gap (''pseudogap'') $E_{JT}$ and a $T$-dependent superconducting gap 
$\Delta _{s}(T)$ \cite{Demsar,Norman,Claeson} for which $\Delta
_{s}(0)\lesssim E_{JT}.$

The proposed model suggests a simple explanation why $T_{c}$ {\em decreases}
in the overdoped regime. With increased doping, the stripe length $l_{s}$
increases leading to increased overall metallicity, while at the same time 
{\em the number of pairs decreases}, leading to a decrease in $T_{c}$
according to the formula given by Eq. (8). Eventually in the metallic,
nonsuperconducting phase, $l_{c}\lesssim l_{p}$ and the material becomes a
homogeneous metal with no pairs and hence the phase no longer supports
high-temperature superconductivity.

Above $T^{\ast }$ the crossover from the underdoped to the overdoped phase
manifests itself in a change of non-degenerate to degenerate statistics as
indicated by region 1 and region 2 respectively in the phase diagram in
Fig.4. In principle they should be distinguishable from the temperature
dependence of the susceptibility for example, which should be Curie-like in
region 1 and Pauli-like in region 2, particularly at low temperatures (see
Figure 4). In contrast, the crossover from region 1 to 3 (Fig. 4) is
governed by excitations across the pseudogap with the temperature dependence
of the susceptibility given by $\chi (T)\propto 1/T^{\alpha }\exp
[E_{JT}/kT] $ \cite{AKM,Mullersusc,NMR}.

\section{Discussion}

An important issue which needs to be discussed is the effective mass of the
non-local JT pairs coupled by $k\neq 0$ wavector phonons. Let us consider -
for simplicity - only the most experimentally relevant interaction for $%
\Sigma $ -point coupling, i.e. the ($k_{x}^{2}-k_{y}^{2})$ term of the
Hamiltonian Eq.(5). In this case we can apply the Lang-Firsov\cite
{Lang-Firsov} transformation which will give an appropriate estimate of the
particle mass \cite{AKR}. In that case, the effective mass renormalization
is exponential:

\begin{equation}
\frac{m^{\ast }}{m_{0}}=\exp {(g_{eff}^{2})}
\end{equation}
where $m_{0}$ is the bare electron mass, and 
\begin{equation}
g_{eff}^{2}=\frac{1}{(2\pi )^{2}}\int d^{2}kg_{k}^{2}[1-\cos {(}ka{)}].
\end{equation}
For simplicity, here the integration is carried over the Cu-O$_{2}$ plane,
so $k$ refers to in-plane momentum. Assuming that the main contribution
comes from $k\simeq k_{0}$, ignoring the effect of $\gamma $ and
integrating, we obtain: 
\begin{equation}
g_{eff}^{2}=g^{2}k_{0}^{4}[1-\cos {(}k{_{0}}a{)}]/8\pi
\end{equation}
This formula can be rewritten in terms of the ground state energy of a
single polaron as: 
\begin{equation}
g_{eff}^{2}=\frac{E_{p}[1-\cos {(}k{{_{0}}}a{)}]}{2\omega }
\end{equation}
where the polaron binding energy $E_{p}=g^{2}k_{0}^{4}\omega /4\pi $. When
compared with the similar expression for the effective mass in the Holstein
model which has no $k$-depedence, we find that the effective mass exponent
is a factor 2 smaller than the corresponding expression in the Holstein
(bi)polaron\cite{Bonca}. If $k_{0}<\pi /2a,$ the effective mass becomes even
smaller, reflecting the fact that for forward scattering the electron-phonon
interaction does not increase the mass strongly. Indeed for $%
k_{0}\rightarrow 0$, the effective mass approaches the bare electron mass $%
m^{\ast }\rightarrow $ $m_{0}$. This effect is similar to that discussed by
Alexandrov for the case of the Froehlich interaction\cite{froelich}.
However, note that in this case the interaction is weak and there is no pair
binding at all for $k=0$, which means that it is not relevant if we are
considering pairing, but {\it is} relevant if we consider single-electron
transport in the normal state. On the other hand if $k_{0}>\pi /2a,$ the
mass enhancement becomes more pronounced because of strong backscattering,
and so at the zone boundary, corresponding to the special points $X$ and $M$
in the BZ, we expect a very large coupling and a strongly enhanced pair
mass. This situation would be relevant to a zone-doubling (for the $M$%
-point) or quadrupling (for the $X$-point) charge density wave formation
and/or the formation of long-range order associated with a structural phase
transition. The case {\em relevant for pairing} is of course intermediate,
as indicated by the wavevector $k_{0}$ in the neutron experiments.

As already discussed, according to the neutron data the interaction in the
cuprates appears to take place over a large range of wavevectors $\gamma $
centered near $k_{0}\sim 1/l_{p}$. An interesting case arises at the 1/8
doping level, where the interparticle distance $l=\sqrt{8}a$. If $l$
corresponds exactly to $l=2\pi /k_{0}$ we expect to observe a CDW\ with a
periodicity given by $k_{0}$. (Note that this is different to the simpler
case of a zone-boundary CDW\ discussed in the previous paragraph.)

In the underdoped state the JT model is different from the bipolaronic (BP)
models\cite{Alexandrov,Bersukher} and other intersite models\cite
{Chakraverty} primarily with regard to the detailed mechanism of bipolaron
formation. Whereas the standard bipolaron model usually refers to
quantum-chemical calculations\cite{Chakraverty} and does not necessarily
involve a particular JT mode, nor a specific local symmetry change upon
pairing, the present intersite JT \ pairing model does so, and implies a
very specific Hamiltonian (Eq. (5)) which is based on the symmetry analysis
of experimentally determined local distortions. Eq. (2) at first sight has
some common features with the phenomenology of the charge-density wave (CDW)
scenario \cite{DiCastro}. The present model offers a microscopic description
for the origin of this interaction as arising from JT-coupling between a $%
k\neq 0$ mode and degenerate electronic states.

The proposed scenario suggests the coexistence of Fermionic excitations in
stripes and Bosons (pairs) over the entire phase diagram in different
proportion determined by thermodynamic equilibrium. This appears to be born
out by the susceptibility data\cite{Mullersusc} and the 2-component
interpretation of the optical conductivity \cite{Timusk,MMM} amongst others 
\cite{review}.

It can also be shown to be consistent with the temperature and doping
dependence of angle-resolved photoemission spectra. A pairing JT deformation
at the $\Sigma $ point leads to objects which have finite dimensions along
the $a$ or $b$ crystal axes. We therefore expect to observe features
associated with these objects in $k$-space along the $\Sigma $ direction
(i.e. along $\Gamma -M)$ and the appearance of a ''pseudogap'' in\ the ARPES
spectra. The range of wavevectors where such a ''pseudogap'' appears is
given by $\Delta k\sim \gamma $ from Eq. (1). The metallic stripes on the
other hand, in which Fermionic excitations exist in the normal state, above $%
T_{c}$ we expect to observe a band which crosses the Fermi-level along the $%
\Sigma $ direction. Importantly, with increased carrier concentration, the
increased {\em coupling between pairs and stripes} leads to increased 2D
order, progressively extending the Fermi surface in the overdoped state.
Clearly, the temperature-dependent superconducting gap $\Delta _{s}(T)$
which forms in the stripes will appear in the same regions in $k$-space as
the Fermionic band. If we assume that the model can be extended to Bi$_{2}$Sr%
$_{2}$CaCu$_{2}$O$_{8+\delta }$, the coexistence of a $T$-dependent
''superconducting'' gap and a ''pseudogap'' along $\Gamma -M$ \ (i.e. the $%
\Sigma $ direction : see Fig. 2), and especially the apparent ''destruction
of the Fermi surface'' with underdoping \cite{Norman} can be understood to
be consequences of the Hamiltonian (5).

Reconciling the slight differences in the interpretation of the observed
lattice distortions in ESR, EXAFS and inelastic neutron scattering, the $%
\Sigma -$ point symmetry analysis of ionic displacements in La$_{2-x}$Sr$%
_{x} $CuO$_{4}$ shows that the distortion of $\tau _{7}$ symmetry at the
zone boundary which was invoked to explain the ESR\cite{Kochelaev} and EXAFS 
\cite{Bianconi} (Figure 6) is in fact the zone-boundary (i.e. short-range)
equivalent to the $\tau _{1}$ distortion occuring over a more extended
length scale along the $\Sigma $ direction in the BZ and the experiments may
be detecting the same mode described by Eq. (5).

We end the discussion by noting that the choice of $k_{0}$ made on the basis
of neutron data also determines the symmetry of the pairing channel in Eq.5.
The first term is isotropic ($s$-wave) while the second one has $d$-wave
symmetry along the Cu-O bond axes. The relative strengths of the terms are
of course to be determined by experiments, but the large phonon anomaly at
the $\Sigma $ point in the inelastic neutron data clearly emphasizes the $d$%
-wave component.

\section{Conclusion}

The main aim of the present paper is to identify an interaction which can
lead to pairing in La$_{2-x}$Sr$_{x}$CuO$_{4}$ on the basis of a symmetry
analysis of the experimentally observed anomalies in the $k\neq 0$ phonon
spectrum. It essentially describes the interaction which causes the
microscopic inhomogeneities observed in experiments. The rest of the paper
is devoted to a discussion of the implications for superconductivity and the
phase diagram. The non-local Jahn-Teller pairing interaction which couples $%
\tau _{1}$ modes at the $\Sigma $ point with degenerate in-plane O $p_{x}$
and $p_{y}$ states is in spirit, if not in detail similar to the motivation
described in the original paper on La$_{2-x}$Ba$_{x}$CuO$_{4}$ by Bednorz
and Muller\cite{BednorzMuller}. The pseudogap in the normal state results
from pair density fluctuations and the temperature $T^{\ast }$ represents an
energy scale for the pairing $kT^{\ast }\sim E_{JT}$ $\simeq 32$ meV for La$%
_{2-x}$Sr$_{x}$CuO$_{4}$. The model naturally leads to the formation of
stripes and the cross-over from a predominantly paired (Bosonic) normal
state to a mixed fermion-boson system in the overdoped region. A
straightforward and appealing way to explain the doping dependence of $T_{c}$
in the overdoped regime by Eq. (8) arises from the fact that at higher
doping levels the average stripe lengths increase and thus {\em the number
of pairs is reduced,} thus reducing $T_{c}$. Apart from giving rise to a
rather simple phase diagram which is consistent with experimental
observations, the model also answers the question why superconductivity
often appears near an orthorhombic phase of the material. However, because
the pairs are dynamic and incommensurate, the locally orthorhombic phase
associated with the JT-pair cannot be easily detected by time- and
spatially- averaging experimental techniques, and one does {\em not} expect
to observe a static orthorhombic phase below $T^{\ast }$. On the other hand,
the model can explain well the inelastic neutron scattering, neutron PDF,
EXAFS, ARPES, susceptibility and ESR, as well others\cite{Loram,Hackl,NMR}
which we have not discussed here.

While here we have mainly focussed on La$_{2-x}$Sr$_{x}$CuO$_{4}$, we note
that similar large $k\neq 0$ lattice distortions have been reported in YBa$%
_{2}$Cu$_{3}$O$_{7-\delta }$\cite{YBCO} and we expect a similar mechanism to
work there also, as well as the other cuprates and oxides in general where
mesoscopic inhomogeneities are observed. We have also omitted a discussion
of the spin coupling associated with the local pairs given by the last term
in Eq. (5), but mention only that in contrast to the Holstein model, the
present Hamiltonian allows the formation of spin singlet {\em or }triplet
pairs\cite{NMR}. Finally we might add as a general comment that a short
superconducting coherence length of the order of the inter-carrier spacing
may be an indication that carriers are paired by a finite-wavevector JT
instability forming non-local pairs. Apart from the cuprates, alkali doped
fullerenes might be an example of such a case.

\section{Acknowledgments}

We wish to acknowledge very useful and encouraging discussions with
K.A.M\"{u}ller, V.Kresin, A.S.Alexandrov and T.Mertelj for important
comments.

\section{Figures}

Figure 1. The distortion in the CuO plane corresponding to the anomalous
mode observed in inelastic neutron scattering in La$_{2-x}$Sr$_{x}$CuO$_{4}$ 
\cite{Egami,Mook}. The O displacements are those of the $\tau _{1}$ mode
shown in Figure 3 and in general have different phase.

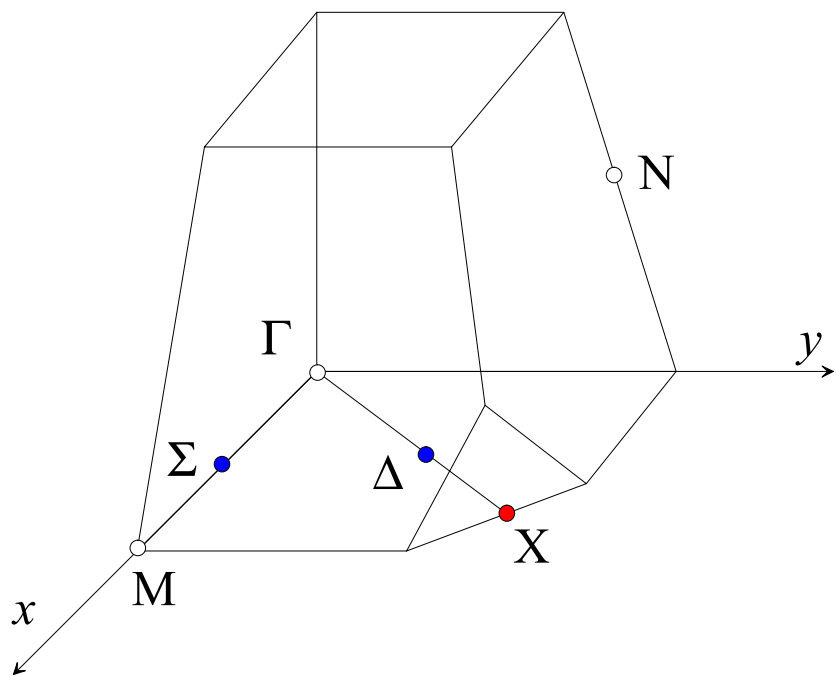
Figure 2. The Brillouin zone (BZ) of La$_{2-x}$Sr$_{x}$CuO$_{4}$
corresponding to the tetragonal phase with point group $D_{4h}.$

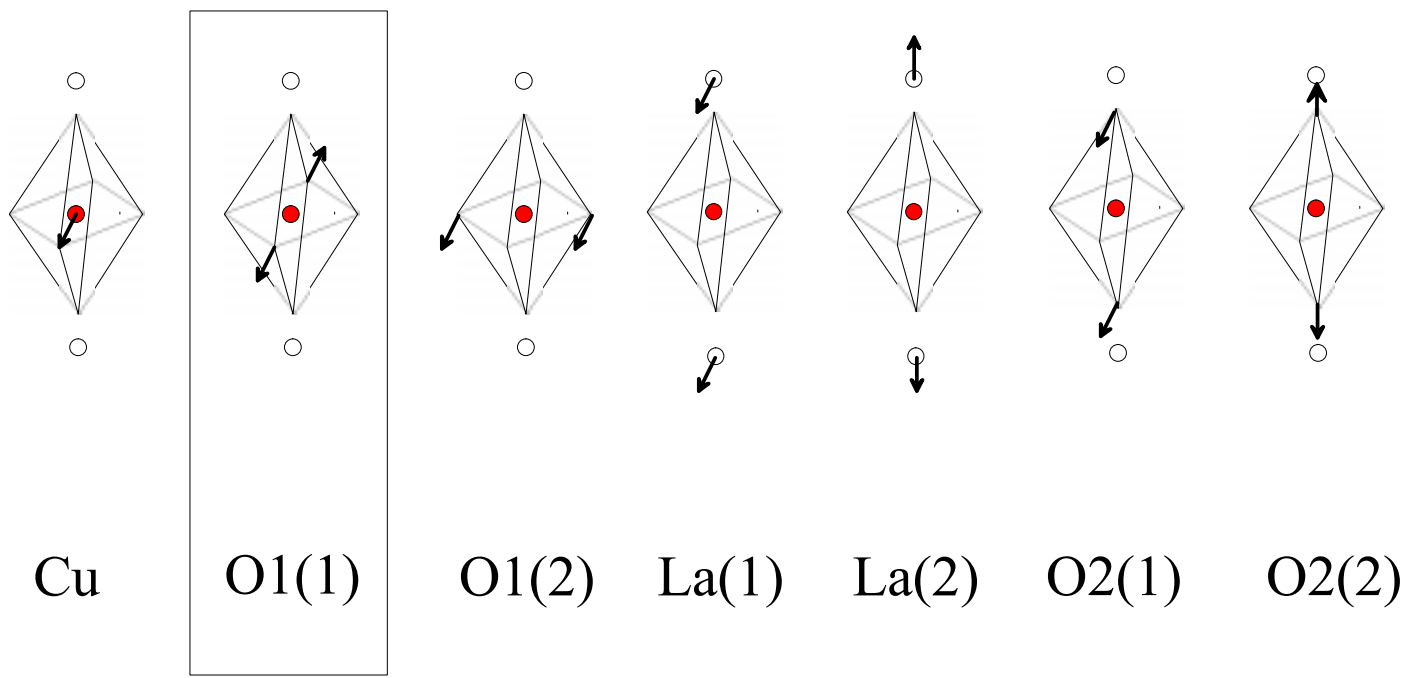
Figure 3. The ionic displacements in La$_{2-x}$Sr$_{x}$CuO$_{4}$
corresponding to $\tau _{1}$ symmetry of the little group at the $\Sigma -$%
point in the BZ. The mode observed in neutron scattering corresponds to the
O1(1) displacements.

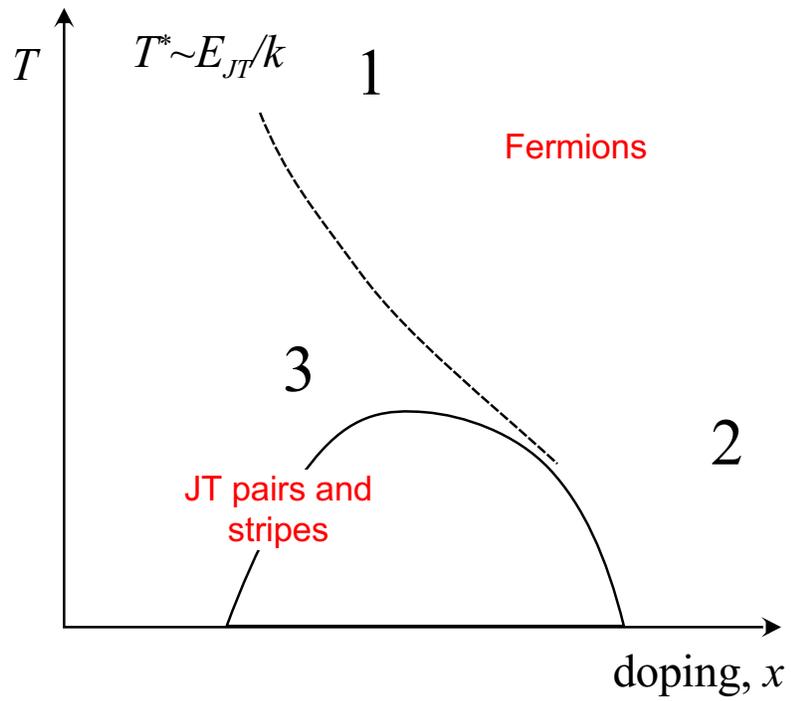
Figure 4. A schematic phase diagram suggested on the basis of the proposed
model for the cuprates. The dashed line indicates the temperature $T^{\ast }$
where $kT^{\ast }\simeq E_{JT}$. The solid line indicates the temperature $%
T_{c}$ of the onset of macroscopic phase coherence and is given by Eq. (7).

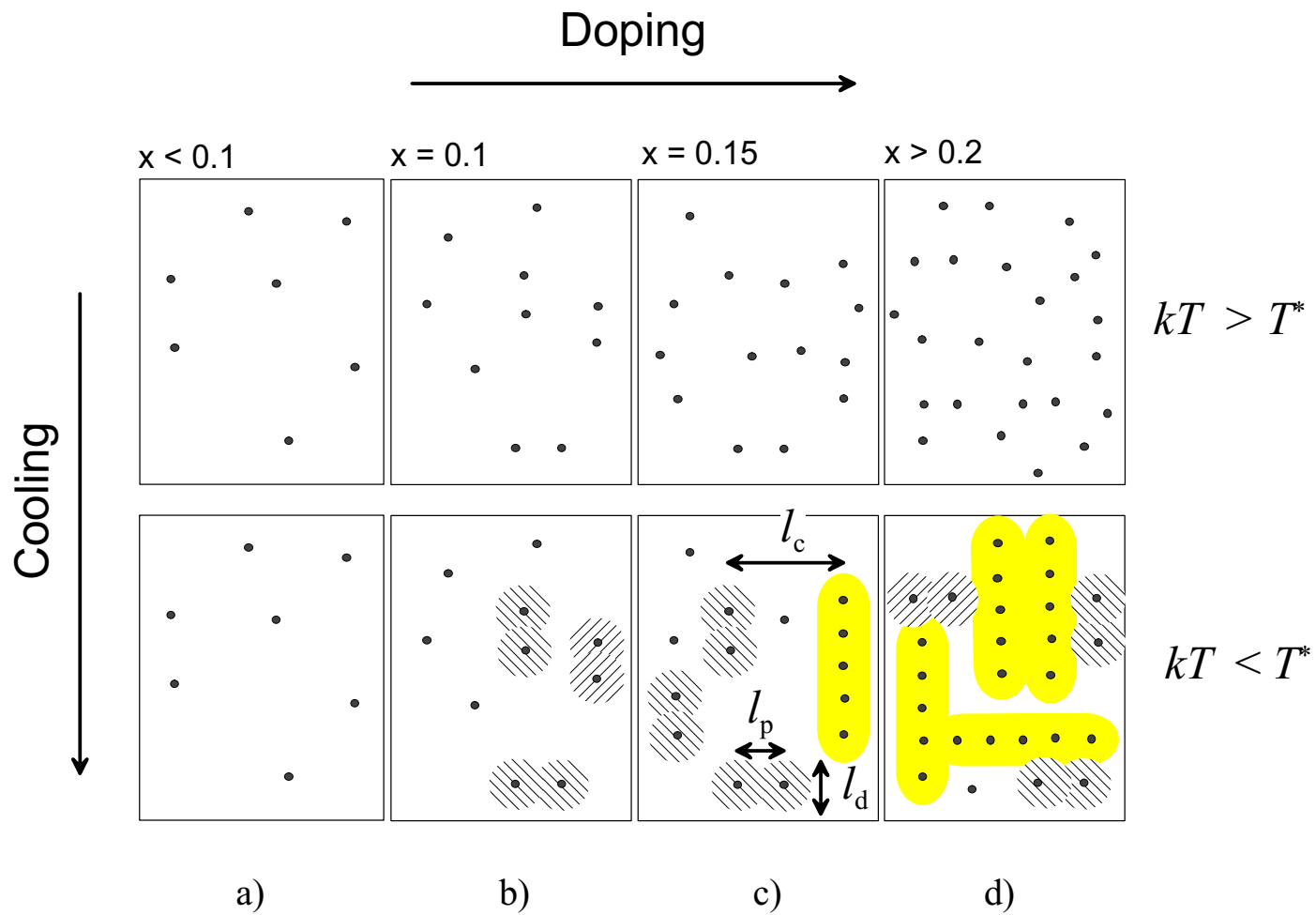
Figure 5. Real-space schematic diagram representing approximately 100 unit
cells ($\approx 4\xi ^{2}$) in the Cu-O plane at different doping levels: a)
for $T>T^{\ast }$ the carriers are unbound single particles (region 1 in the
phase diagram in Figure 4), b) for $T<T^{\ast }$in the underdoped state
(region 3 in Fig. 4) pairs and unbound particles co-exist with few stripes.
For $T<T^{\ast }$\ near optimum doping and in the overdoped state (c) and d)
respectively) pairs coexist with unbound particles and stripes.

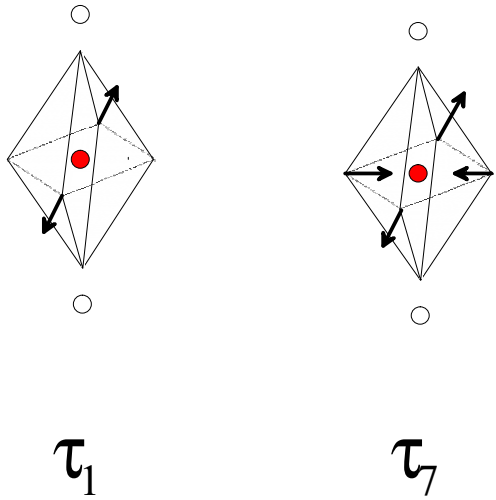
Figure 6. A superposition of two $\tau _{1}$ modes (observed in neutron
scattering by Egami\cite{Egami})\ with orthogonal $k$-vectors at the $\Sigma 
$ point has the same displacements as the $\tau _{7}$ mode at the zone
boundary observed in ESR \cite{Kochelaev}.


\begin{references}
\bibitem{direction}  Note that a distortion along 45$^{\circ }$ to the Cu-O
bonds ($\Delta -$point$)$ is also allowed, but is not discussed further
because the experiments suggest that the $\Sigma $-point is more relevant.
\bibitem{Bonca}  Similar results were obtained by numerical exact
calculations with the Hubbard model, where for intersite pairs, the
effective mass is of the order of the single polaron mass. see J.Bonca and
S.Trugman (to be published, 2000)
\end{references}
\end{document}